\documentstyle[prl,aps,preprint,eqsecnum]{revtex}
\begin{document}
\draft
\title{Irreducible decompositions in metric-affine gravity models}

\author{Yu.N.\ Obukhov\footnote{Permanent address: Department of Theoretical
Physics, Moscow State University, 117234 Moscow, Russia},
E.J.\ Vlachynsky\footnote{Permanent address: Department of Mathematics, 
University of Newcastle, Newcastle, NSW 2308, Australia},
 W.\ Esser, and F.W.\ Hehl}
\address
{Institute for Theoretical Physics,
University of Cologne\\D-50923 K{\"o}ln, Germany}

\maketitle

\begin{abstract}
The irreducible decomposition technique is applied to the study of 
classical  models of metric-affine gravity (MAG). The dynamics of the
gravitational field is described by a 12-parameter Lagrangian 
encompassing a Hilbert-Einstein term, torsion and nonmetricity
square terms, and one quadratic curvature piece that is built up from 
Weyl's segmental curvature. Matter is represented by a hyperfluid, a 
continuous medium the elements of which possess classical momentum and 
hypermomentum. With the help of irreducible decompositions, we are able 
to express torsion and traceless nonmetricity explicitly in terms of the 
spin and the shear current of the hyperfluid. Thereby the field equations 
reduce to an effective Einstein theory describing a metric coupled to the 
Weyl 1-form (a Proca-type vector field) and to a spin fluid. We 
demonstrate that a triplet of torsion and nonmetricity 1-forms describes 
the general and unique vacuum solution of the field equations of MAG. 
Finally, we study homogeneous cosmologies with an hyperfluid. We find that 
the hypermomentum affects significantly the cosmological evolution at very 
early stages. However, unlike spin, shear does not prevent the formation of a 
cosmological singularity.
\end{abstract}
\pacs{PACS no.: 04.50.+h; 04.20.Jb; 03.50.Kk}

\section{Introduction}
  
Within the framework of the gauge approach to gravity (see, e.g.,
\cite{PR}) the {\it kinematic} scheme of the metric-affine theory is well 
understood at present. However, the {\it dynamic} aspects of metric-affine 
gravity (MAG) have been rather poorly studied up to now. The choice of the 
basic Lagrangian of the theory remains an open problem, and this, in turn, 
prevents a detailed analysis of possible physical effects. [An analysis of 
physical observations in MAG without specializing to a particular 
Lagrangian was made in \cite{coley}.] As a first step, 
one can use a correspondence principle. It is well known that Einstein's 
general relativity theory is satisfactorily supported by experimental tests 
on the macroscopic level. Thus, whereas the gravitational gauge models 
provide an alternative description of gravitational physics in the microworld, 
it is natural to require their correspondence with general relativity at 
large distances. Unfortunately, direct generalization of the standard 
Hilbert-Einstein Lagrangian yields an unphysical MAG model which is 
projectively invariant and, accordingly, imposes unphysical constraints 
on the matter sources. 

Another essential difficulty in the development of a dynamical scheme of 
MAG was, until recently, the lack of self-consistent models which
describe physical (quantum, semiclassical, or classical) sources of MAG 
possessing mass or energy-momentum and hypermomentum. The reader may 
consult \cite{PR,nee} which give a modern presentation of the so called 
manifield and world spinor approach based on the theory of infinite
dimensional representations of the affine and linear groups. However, 
the main achievements there are again of kinematic nature, and no 
dynamical model for manifields and world spinors is available. 

Recently there has been some progress both in the development of the
simplest viable metric-affine Lagrangians that generalize the 
Hilbert-Einstein model and in the establishment of a variational theory 
of a {\it hyperfluid} which seems to represent a reasonable classical model 
of a continuous medium with energy-momentum and hypermomentum. In the papers 
\cite{TW,gh} it was proposed to take as the gravitational Lagrangian the sum 
of the (generalized) Hilbert-Einstein term and the square of the segmental
curvature (thus reviving the old proposals of \cite{Palatini,Pono}). Further 
extensions of this model, which include the quadratic invariants of torsion 
and nonmetricity, were investigated (in vacuum) in 
\cite{Tres1,Tres2,magexact,magkerr,tuck}. A hyperfluid model was developed 
in \cite{flu} along the lines of the Weyssenhoff approach to spin fluids 
which now reappear as a particular case of the hyperfluid. [Note that a 
different variational model of a fluid with hypermomentum was suggested in 
\cite{smal}].

Relativistic fluid dynamics covers a vast field of research in gravitation,
cosmology, and particle physics. Relativistic fluid models are working tools 
in high-energy plasma astrophysics and in nuclear physics (where non-ideal 
fluids are extremely successfully applied to the description of heavy ion 
reactions), see, e.g., \cite{fluid}. In cosmology, hydrodynamical description 
of matter is standard both for the early and for the later stages of the 
evolution of the universe \cite{ellis}. Spin fluids are used for the
consistent statistical treatment of a medium the elements of which are 
particles with intrinsic angular momentum \cite{israel} (cosmological ``soup''
of fundamental particles in the early universe or a fluid of spinning galaxies,
clusters of galaxies, turbulent eddies during the later times). In 
Poincar\'e gauge gravity, the Weyssenhoff spin fluid \cite{spinflu} provides
an adequate description of a continuous medium with spin degrees of freedom.
Spin of matter sources proves to be significant in the Einstein-Cartan theory,
where the cosmological singularity can be avoided due to effective repulsion
of spinning particles \cite{tra}. It is worthwhile to stress that the spin
fluid can be consistently derived from the quantum theory of Dirac particles.
[One may ask though: How can this be reconciled with the studies \cite{kerlick}
for the Einstein-Cartan-Dirac cosmology where a singularity is not avoided? 
The answer is that it is misleading to compare a spin fluid cosmology (i.e., 
a cosmology of {\it ensemble of large number} of gravitating particles with 
spin) with a clearly unphysical classical Dirac field ``cosmology''(i.e., a
``cosmology'' of {\it one} gravitating particle with spin). The correct 
comparison can only be made with the Einstein-Cartan-Dirac cosmology in which 
the energy-momentum and spin currents are obtained as macroscopic averages 
from the quantum density operators. This was done, e.g., in \cite{bau} with 
the help of the relativistic Wigner function formalism, and the effective 
repulsion was confirmed]. 

In the framework of MAG, to the best of our knowledge, the hyperfluid 
represents the only available self-consistent dynamical model of matter with 
nontrivial hypermomentum. It generalizes in a natural way the Weyssenhoff 
spin fluid by including additional degrees of freedom (dilation and shear 
densities). At the same time, further study is certainly needed for 
establishing the {\it fundamental} theory of matter with hypermomentum. The 
manifields \cite{PR,nee} seem to be a step in the right direction, but 
unfortunately no dynamical scheme is known for them (i.e., no Lagrangian, no 
equations of motion, no precise form of the Noether currents). It is even 
unclear how the standard Dirac fermion matter can be recovered from the 
manifields when shear and dilation charges vanish. We are convinced though, 
that even after the fundamental theory of matter with hypermomentum is 
completed, the hyperfluid model will remain a tool useful for practical 
applications (like the relativistic fluid models are amazingly handy in 
nuclear physics for calculations on heavy ion reactions, despite the fact 
that a fundamental Dirac theory is always also available \cite{fluid}).

In this paper we will study the classical dynamics of metric-affine 
gravitational fields for the general MAG Lagrangian which includes the 
Hilbert-Einstein term, the segmental curvature square term (of Weyl), and
all possible quadratic torsion and nonmetricity contractions. The hyperfluid
provides nontrivial energy-momentum and hypermomentum currents which
describe classical matter sources in the MAG field equations. We 
demonstrate the exceptional effectiveness of the technique of irreducible 
decompositions applied to post-Riemannian geometrical objects. In particular,
we show that (i) the separation of Riemannian and post-Riemannian
structures and (ii) the subsequent decomposition of the latter into 
irreducible pieces, leads to the solution of the coupled MAG field equations 
with respect to torsion and nonmetricity. As a result of this process, we are 
left with an effective Einsteinian gravitational field equation for the 
metric which is a direct generalization of the effective equations 
arising in the Einstein-Cartan theory (cf. \cite{tuck2}). Specializing our
results to the vacuum case, we are able to complete the study of the 
ansatz of the so-called {\it 1-form triplet}, which underlies the results of 
\cite{magexact,magkerr,tuck}, by demonstrating its uniqueness. Namely, for
generic MAG models, the solution with a 1-form  triplet is not only the most 
general solution of the {\it second} field equation (for terminology, see 
\cite{PR}) but it is also unique. 

Our basic notations and conventions are those of \cite{PR}, in particular
the signature of the metric is assumed to be $(-,+,+,+)$. 

\section{Preliminaries: metric-affine geometry and basic decompositions}

In this section we recall some basic facts concerning metric-affine geometry
in four dimensions. For a more detailed discussion in arbitrary dimensions
see \cite{PR}. The metric-affine spacetime is described by the metric
$g_{\alpha\beta}$, the coframe 1-forms $\vartheta^{\alpha}$, and the linear
connection 1-forms $\Gamma_{\beta}{}^{\alpha}$. These are interpreted as
the generalized gauge potentials, while the corresponding field strengths
are the nonmetricity 1-form $Q_{\alpha\beta}=-Dg_{\alpha\beta}$ and the
2-forms of torsion $T^{\alpha}=D\vartheta^{\alpha}$ and curvature 
$R_{\beta}{}^{\alpha}=d\Gamma_{\beta}{}^{\alpha} + \Gamma_{\gamma}{}^{\alpha}
\wedge\Gamma_{\beta}{}^{\gamma}$. The metric-affine geometry reduces to a 
purely Riemannian one as soon as torsion and nonmetricity both vanish. 
It is well known that for every metric $g_{\alpha\beta}$ there exists 
a unique torsion-free and metric-compatible connection represented by the 
Christoffel symbols. We will denote this Riemannian connection by $\widetilde{
\Gamma}_{\beta}{}^{\alpha}$, and hereafter the tilde will denote purely
Riemannian geometrical objects and covariant differentials constructed from
them.  

The general affine connection can always be decomposed 
into Riemannian and post-Riemannian parts,
\begin{equation}
\Gamma_{\beta}{}^{\alpha}=\widetilde{\Gamma}_{\beta}{}^{\alpha} +
N_{\beta}{}^{\alpha},\label{decom}
\end{equation}
where the {\it distortion} 1-form $N_{\alpha\beta}$ can be 
expressed in terms of torsion and nonmetricity:
\begin{equation}
N_{\alpha\beta}=-e_{[\alpha}\rfloor T_{\beta]} + 
{1\over 2}(e_{\alpha}\rfloor e_{\beta}\rfloor T_{\gamma})\vartheta^{\gamma}
+ (e_{[\alpha}\rfloor Q_{\beta]\gamma})\vartheta^{\gamma} 
+{1\over 2}Q_{\alpha\beta}.\label{N}
\end{equation}
Using (\ref{decom}), it is possible to split all quantities in the 
metric-affine theory into Riemannian and post-Riemannian pieces (for curvature 
this reads, e.g.: $R_{\beta}{}^{\alpha}=\widetilde{
R}_{\beta}{}^{\alpha} + \widetilde{D}N_{\beta}{}^{\alpha} + 
N_{\gamma}{}^{\alpha}\wedge N_{\beta}{}^{\gamma}$).

Let us turn to the description of {\it irreducible decompositions} of 
post-Riemannian geometrical objects. However, as the complete decomposition 
of curvature will not be needed in this paper, the interested reader is 
referred to \cite{PR} instead. For us the most important decompositions
will be those of torsion and nonmetricity since they, as we will see 
later, provide a pattern for the decomposition of the gravitational gauge 
field momenta.

\subsection{Decomposition of torsion}

The torsion 2-form can be decomposed into three irreducible pieces, 
$T^{\alpha}={}^{(1)}T^{\alpha} + {}^{(2)}T^{\alpha} + {}^{(3)}T^{\alpha}$,
where
\begin{eqnarray}
{}^{(2)}T^{\alpha}&:=&{1\over 3}\vartheta^{\alpha}\wedge T,\label{T2}\\
{}^{(3)}T^{\alpha}&:=&-\,{1\over 3}{}^*(\vartheta^{\alpha}\wedge P),
\label{T3}\\
{}^{(1)}T^{\alpha}&:=&T^{\alpha}-{}^{(2)}T^{\alpha} - {}^{(3)}T^{\alpha}.
\label{T1}
\end{eqnarray}
The torsion trace (covector) and pseudotrace (axial covector) 1-forms are 
defined, respectively, by
\begin{equation}
T:=e_{\alpha}\rfloor T^{\alpha},\quad\quad 
P:={}^*(T^{\alpha}\wedge\vartheta_{\alpha}).
\end{equation}

\subsection{Decomposition of nonmetricity}

The nonmetricity 1-form can be decomposed into four irreducible pieces, 
$Q_{\alpha\beta}={}^{(1)}Q_{\alpha\beta}+{}^{(2)}Q_{\alpha\beta}+
{}^{(3)}Q_{\alpha\beta}+{}^{(4)}Q_{\alpha\beta}$, with 
\begin{eqnarray}
{}^{(2)}Q_{\alpha\beta}&:=&{2\over 3}\,{}^*\!(\vartheta_{(\alpha}\wedge
\Omega_{\beta)}),\label{Q2}\\
{}^{(3)}Q_{\alpha\beta}&:=&{4\over 9}
\left(\vartheta_{(\alpha}e_{\beta)}
\rfloor\Lambda - {1\over 4}g_{\alpha\beta}\Lambda\right),\label{Q3}\\
{}^{(4)}Q_{\alpha\beta}&:=&g_{\alpha\beta}Q,\label{Q4}\\
{}^{(1)}Q_{\alpha\beta}&:=&Q_{\alpha\beta}-{}^{(2)}Q_{\alpha\beta}-
{}^{(3)}Q_{\alpha\beta}-{}^{(4)}Q_{\alpha\beta}.\label{Q1}
\end{eqnarray}
Here the shear covector part and the Weyl covector are, respectively,
\begin{equation}
\Lambda:=\vartheta^{\alpha}e^{\beta}\rfloor
{\nearrow\!\!\!\!\!\!\!Q}_{\alpha\beta},\quad\quad
Q:={1\over 4}g^{\alpha\beta}Q_{\alpha\beta},
\end{equation}
where ${\nearrow\!\!\!\!\!\!\!Q}_{\alpha\beta}=Q_{\alpha\beta}-
Qg_{\alpha\beta}$ is the traceless piece of the nonmetricity. 

It seems worthwhile to notice that the 2-form $\Omega^{\alpha}$, defined by 
$\Omega_{\alpha}:=\Theta_{\alpha} - {1\over 3}e_{\alpha}\rfloor
(\vartheta^{\beta}\wedge\Theta_{\beta})$ with $\Theta_{\alpha}:=
{}^*({\nearrow\!\!\!\!\!\!\!Q}_{\alpha\beta}\wedge\vartheta^{\beta})$,
which describes ${}^{(2)}Q_{\alpha\beta}$, has precisely the same symmetry 
properties as the 2-form ${}^{(1)}T^{\alpha}$. In particular, we can prove 
that $e_{\alpha}\rfloor\Omega^{\alpha}=0$ and 
$\vartheta_{\alpha}\wedge\Omega^{\alpha}=0$.

\subsection{Decomposition of the distortion}

Substituting (\ref{T2})-(\ref{T1}) and (\ref{Q2})-(\ref{Q1}) into (\ref{N}),
we find the following general decomposition of the distortion 1-form:
\begin{equation}
N_{\alpha\beta}={1\over 2}\left\{Q_{\alpha\beta} - {2\over 3}
\vartheta_{[\alpha}e_{\beta]}\rfloor (3Q - \Lambda + 2T) -
2e_{[\alpha}\rfloor\left({}^*\Omega_{\beta]} + 2\,{}^{(1)}T_{\beta]}\right)
- \left(e_{\alpha}\rfloor e_{\beta}\rfloor {}^{(3)}T_{\gamma}\right)
\vartheta^{\gamma}\right\}.\label{N1}
\end{equation}
The 1st irreducible piece of the torsion and the 2-form $\Omega_{\alpha}$, 
which represents the 2nd irreducible piece of nonmetricity, appear as a 
linear combination in (\ref{N1}). This general formula proves to be 
extremely useful in the analysis of the field equations of MAG.

\section{A model for MAG}

In this and subsequent sections we will widely use, along with the coframes
$\vartheta^{\alpha}$, the so called $\eta$-basis of the dual coframes. 
Namely, we define \cite{PR} the Hodge dual such that $\eta:={}^*1$ is the 
volume 4-form. Furthermore $\eta_{\alpha}:=e_{\alpha}\rfloor\eta =
{}^*\vartheta_{\alpha}$, $\eta_{\alpha\beta}:= e_{\beta}\rfloor
\eta_{\alpha}={}^*(\vartheta_{\alpha}\wedge\vartheta_{\beta})$, $\eta_{\alpha
\beta\gamma}:=e_{\gamma}\rfloor\eta_{\alpha\beta}$, $\eta_{\alpha\beta\gamma
\delta}:=e_{\delta}\rfloor\eta_{\alpha\beta\gamma}$. The last expression is 
thus the totally antisymmetric Levi-Civita tensor. 

\subsection{Gravitational Lagrangian}

Direct generalization of the Hilbert-Einstein Lagrangian $R_{\alpha\beta}
\wedge\eta^{\alpha\beta}$ to metric-affine gravity yields an unphysical
model which is invariant under projective transformations of the
connection. 

Consequently, we turn our attention to a model described by a Lagrangian 
which generalizes the models studied recently in 
\cite{TW,magexact,magkerr,tuck},
\begin{eqnarray} 
V_{\rm MAG}&=&
\frac{1}{2\kappa}\,\left[-a_0\,R^{\alpha\beta}\wedge\eta_{\alpha\beta}
-2\lambda\,\eta+ T^\alpha\wedge{}^*\!\left(\sum_{I=1}^{3}a_{I}\,^{(I)}
T_\alpha\right)\right.\nonumber\\
&&+ 2\left(\sum_{I=2}^{4}c_{I}\,^{(I)}Q_{\alpha\beta}\right)
\wedge\vartheta^\alpha\wedge{}^*\!\, T^\beta + Q_{\alpha\beta}
\wedge{}^*\!\left(\sum_{I=1}^{4}b_{I}\,^{(I)}Q^{\alpha\beta}\right)
\nonumber\\
&&+ b_5({}^{(3)}Q_{\alpha\gamma}\wedge\vartheta^{\alpha})\wedge
{}^*({}^{(4)}Q^{\beta\gamma}\wedge\vartheta_{\beta})\Bigg] - \frac{1}{2}
z_4\,R^{\alpha\beta}\wedge{}^*{}^{(4)}Z_{\alpha\beta}.\label{lagr}
\end{eqnarray} 
Here, the coupling constants $a_0,...,a_3,c_2,c_3,c_4, b_1,...,b_5, z_4$ are 
dimensionless, $\kappa$ is the standard Einstein gravitational constant,
and $\lambda$ is the cosmological constant. The segmental curvature is
denoted by ${}^{(4)}Z_{\alpha\beta}:={1\over 4}g_{\alpha\beta}R_{\gamma}{}
^{\gamma}$; it is a purely {\it post}-Riemannian piece. 

\subsection{Hyperfluid matter}

Let us study the model (\ref{lagr}) with matter represented by a
hyperfluid \cite{flu}. The matter Lagrangian reads
\begin{equation}
L_{\rm mat}={1\over 2}\rho\,\mu^{A}{}_{B}\,b^{B}_{\alpha}\,u\wedge 
Db^{\alpha}_{A} - \varepsilon(\rho, s, \mu^{A}{}_{B})\,\eta + 
L_{\rm constraints},\label{mat}
\end{equation}
where the first two terms on the right-hand side describe the kinetic and 
the internal energy density $\varepsilon$ of the hyperfluid, respectively. 
The latter depends on the particle density $\rho$, the specific entropy $s$, 
and the specific hypermomentum density $\mu^{A}{}_{B}$ (``specific" means 
``per particle"). Here $u$ is the flow 3-form, so that the components of the 
average fluid velocity are given by $u_{\alpha}:=e_{\alpha}\rfloor^\ast u$. 
The first term represents the combined kinetic contribution of the rotational 
and the strain energy of the fluid elements the motion of which is described 
by the angular and strain velocity of a 3-volume spanned by the material 
triad. It is convenient to describe the latter by two variables: a 1-form 
$b^A$ with the components $b^{A}_{\alpha}:= e_{\alpha}\rfloor b^{A}$ and a 
3-form $b_A$ with the components $b_{A}^{\alpha}:={}^\ast(b_{A}\wedge
\vartheta^{\alpha})$. The last term in (\ref{mat}) denotes a set of 
constraints to be added via Lagrange multipliers. We will not display them 
here (see \cite{flu} for a detailed discussion). Let us only mention that 
they include the standard normalization constraint for the velocity 
\begin{equation}
u\wedge{}^{\ast}u =\eta,\label{norm}
\end{equation}
and the law of particle number conservation,
\begin{equation}
d(\rho u)=0.\label{num}
\end{equation}
Variation of the Lagrangian (\ref{mat}) with respect to the matter variables
yields the hypermomentum equation of motion in the form
\begin{equation}
D\Delta^{\alpha}{}_{\beta}= - u^{\alpha}\,u_{\lambda}D\Delta^{\lambda}
{}_{\beta} - u_{\beta}\,u^{\lambda}D\Delta^{\alpha}{}_{\lambda},\label{hy}
\end{equation}
where the hypermomentum current 3-form 
\begin{equation}
\Delta^{\alpha}{}_{\beta}=uJ^{\alpha}{}_{\beta}\label{hcur}
\end{equation}
can be expressed in terms of the hypermomentum density
\begin{equation}
J^{\alpha}{}_{\beta}=-{1\over 2}\rho\,\mu^{A}{}_{B}\,b^{B}_{\beta}\,
b^{\alpha}_{A}.\label{hden}
\end{equation}
By construction, this tensor satisfies the generalized Frenkel conditions
\begin{equation}
J^{\alpha}{}_{\beta}\,u^{\beta}=0,\quad\quad 
J^{\alpha}{}_{\beta}\,u_{\alpha}=0.\label{frenk}
\end{equation}

Variational derivatives of the material Lagrangian $L_{\rm mat}$ with 
respect to coframe $\vartheta^{\alpha}$ and connection $\Gamma_{\beta}
{}^{\alpha}$ 1-forms define the material sources. The canonical hypermomentum 
current 3-form is given by (\ref{hcur}), whereas the canonical energy-momentum 
3-form reads
\begin{equation}
\Sigma_{\alpha}=\varepsilon u u_{\alpha} + p(\eta_{\alpha} + uu_{\alpha}) +
2uu^{\beta}g_{\gamma[\alpha}{\ }^{\ast}(D\Delta^{\gamma}{}_{\beta]}),
\label{en}
\end{equation}
with the pressure defined as usual by $p=\rho{\partial\varepsilon\over
\partial\rho}-\varepsilon$. 

\subsection{Field equations}

The metric-affine field equations are derived from the total
Lagrangian $V_{\rm MAG} + L_{\rm mat}$ by independent variations with
respect to the coframe $\vartheta^{\alpha}$ and connection $\Gamma_{\beta}
{}^{\alpha}$ 1-forms. The corresponding so-called {\it first} and {\it second}
field equations read 
\begin{eqnarray} 
DH_{\alpha}-
E_{\alpha}&=&\Sigma_{\alpha}\,,\label{first}\\ DH^{\alpha}{}_{\beta}-
E^{\alpha}{}_{\beta}&=&\Delta^{\alpha}{}_{\beta}\,.
\label{second}
\end{eqnarray} 
The left hand sides of (\ref{first})--(\ref{second}) are given by
\begin{eqnarray}
M^{\alpha\beta}:=-2{\partial V_{\rm MAG}\over \partial Q_{\alpha\beta}}&=&
-{2\over\kappa}\Bigg[{}^*\! \left(\sum_{I=1}^{4}b_{I}{}^{(I)}
Q^{\alpha\beta}\right) + {1\over 2}b_5\left(\vartheta^{(\alpha}\wedge{}^*
(Q\wedge\vartheta^{\beta)}) - {1\over 4}g^{\alpha\beta}\,{}^*(3Q + \Lambda)
\right)\nonumber\\
&& +\, c_{2}\,\vartheta^{(\alpha}\wedge{}^*\! ^{(1)}T^{\beta)} +
c_{3}\,\vartheta^{(\alpha}\wedge{}^*\! ^{(2)}T^{\beta)} +
{1\over 4}(c_{3}-c_{4})\,g^{\alpha\beta}{}^*\!\,  T\Bigg]\,,\label{M1}\\  
H_{\alpha}:=-{\partial V_{\rm MAG}\over \partial T^{\alpha}} &=& -
  {1\over\kappa}\,
  {}^*\!\left[\left(\sum_{I=1}^{3}a_{I}{}^{(I)}T_{\alpha}\right) +
    \left(\sum_{I=2}^{4}c_{I}{}^{(I)}
      Q_{\alpha\beta}\wedge\vartheta^{\beta}\right)\right],\label{Ha1}\\  
      H^{\alpha}{}_{\beta}:= - {\partial V_{\rm MAG}\over \partial
    R_{\alpha}{}^{\beta}}&=& {a_0\over 2\kappa}\,\eta^{\alpha}{}_{\beta} +
  z_4\,{}^{*}\!\left({}^{(4)}Z^{\alpha}{}_{\beta}\right),\label{Hab1}
\end{eqnarray}
and
\begin{equation}
E_{\alpha}:=e_{\alpha}\rfloor V_{\rm MAG} + (e_{\alpha}\rfloor T^{\beta})
\wedge H_{\beta} + (e_{\alpha}\rfloor R_{\beta}{}^{\gamma})\wedge
H^{\beta}{}_{\gamma} + {1\over 2}(e_{\alpha}\rfloor Q_{\beta\gamma})
M^{\beta\gamma},\label{Ea}
\end{equation}
\begin{equation}
E^{\alpha}{}_{\beta}:= - \vartheta^{\alpha}\wedge H_{\beta} - 
M^{\alpha}{}_{\beta}.\label{Eab}
\end{equation}

We note, see \cite{PR}, that the equation which arise from the variation 
of the Lagrangian with respect to the metric turns out to be redundant.

\section{Special quadratic MAG-Lagrangian}
\medskip

As a preliminary step, let us study the MAG model with the Lagrangian
\begin{eqnarray}
V^{(0)}&=& {a_0\over 2\kappa}\Bigg\{- R_{\alpha\beta}\wedge
\eta^{\alpha\beta} - {}^{(1)}T^{\alpha}\wedge{}^{*(1)}T_{\alpha}
+2{}^{(2)}T^{\alpha}\wedge{}^{*(2)}T_{\alpha} +
{1\over 2}{}^{(3)}T^{\alpha}\wedge{}^{*(3)}T_{\alpha}\nonumber\\
&&+ {}^{(2)}Q_{\alpha\beta}\wedge\vartheta^{\beta}\wedge
{}^{*(1)}T^{\alpha} - 2{}^{(3)}Q_{\alpha\beta}\wedge
\vartheta^{\beta}\wedge{}^{*(2)}T^{\alpha} -2{}^{(4)}Q_{\alpha\beta}
\wedge\vartheta^{\beta}\wedge{}^{*(2)}T^{\alpha}\nonumber\\
&&+ {1\over 4}{}^{(1)}Q_{\alpha\beta}\wedge{}^{*(1)}Q^{\alpha\beta}
-{1\over 2}{}^{(2)}Q_{\alpha\beta}\wedge{}^{*(2)}Q^{\alpha\beta}
-{1\over 8}{}^{(3)}Q_{\alpha\beta}\wedge{}^{*(3)}Q^{\alpha\beta}
\nonumber\\
&&+{3\over 8}{}^{(4)}Q_{\alpha\beta}\wedge{}^{*(4)}Q^{\alpha\beta}
+ ({}^{(3)}Q_{\alpha\gamma}\wedge\vartheta^{\alpha})\wedge
{}^*({}^{(4)}Q^{\beta\gamma}\wedge\vartheta_{\beta})\Bigg\}.\label{V0}
\end{eqnarray}
It can be seen that this is a particular case of (\ref{lagr}) with the 
special values of the coupling constants.

One can verify, by direct calculation, that the gauge field momenta for the
Lagrangian (\ref{V0}) are given by
\begin{equation}
H^{(0)}_{\alpha}:=-{\partial V^{(0)}\over \partial T^{\alpha}}\equiv 
-\,{a_0\over 2\kappa}\,N^{\mu\nu}\wedge\eta_{\alpha\mu\nu},\quad\quad
H^{(0)\alpha}{}_{\beta}:= -{\partial V^{(0)}\over\partial R_{\alpha}{}^{\beta}}
={a_0\over 2\kappa}\,\eta^{\alpha}{}_{\beta},
\end{equation}
and, as a result, it can be straightforwardly proved that 
\begin{equation}
DH^{(0)}_{\alpha}- E^{(0)}_{\alpha}\equiv {a_0\over 2\kappa}\,
\widetilde{R}^{\mu\nu}\wedge\eta_{\alpha\mu\nu},\quad\quad
DH^{(0)\alpha}{}_{\beta}- E^{(0)\alpha}{}_{\beta}\equiv 0,\label{ident}
\end{equation}
where, similarly to (\ref{Ea}) and (\ref{Eab}), 
\begin{equation}
E^{(0)}_{\alpha}:=e_{\alpha}\rfloor V^{(0)} + (e_{\alpha}\rfloor T^{\beta})
\wedge H^{(0)}_{\beta} + (e_{\alpha}\rfloor R_{\beta}{}^{\gamma})\wedge
H^{(0)\beta}{}_{\gamma} + {1\over 2}(e_{\alpha}\rfloor Q_{\beta\gamma})
M^{(0)\beta\gamma},\label{Ea0}
\end{equation}
\begin{equation}
E^{(0)\alpha}{}_{\beta}:= - \vartheta^{\alpha}\wedge H^{(0)}_{\beta} - 
M^{(0)\alpha}{}_{\beta}.\label{Eab0}
\end{equation}

The {\it identities} (\ref{ident}) can be justified by the fact that
\begin{equation}
V^{(0)}\equiv \,{a_0\over 2\kappa}\left\{-\,\widetilde{R}_{\alpha\beta}
\wedge\eta^{\alpha\beta} + d\left[\vartheta^{\alpha}\wedge{}^*\left(
2T_{\alpha} - Q_{\alpha\beta}\wedge\vartheta^{\beta}\right)\right]
\right\}\label{hilb}
\end{equation}
is, up to an exact form, the {\it purely Riemannian} Hilbert-Einstein 
Lagrangian of general relativity theory. However, this observation does not 
provide a rigorous proof of (\ref{ident}), because in the derivation of the 
gauge field equations (\ref{first}) and (\ref{second}), see \cite{PR}, one 
assumes that the gravitational Lagrangian contains frame derivatives, 
$d\vartheta^{\alpha}$, only implicitly in torsion, while (\ref{hilb}) contains 
such terms in the Riemannian connection. Hence, a direct proof is required,
and a rather long calculation involving the irreducible decomposition 
(\ref{N1}) of the distortion 1-form $N_{\alpha\beta}$ demonstrates that 
(\ref{ident}) is true, indeed.

\section{Decomposition of the field equations of MAG}

Let us write the Lagrangian (\ref{lagr}) as
\begin{equation}
V_{\rm MAG}=V^{(0)} + \widehat{V},\label{split}
\end{equation}
where
\begin{eqnarray} 
\widehat{V}&=&
\frac{1}{2\kappa}\,\left[-2\lambda\,\eta 
+ T^\alpha\wedge{}^*\!\left(\sum_{I=1}^{3}\alpha_{I}\,^{(I)}
T_\alpha\right)\right.\nonumber\\
&&+ 2\left(\sum_{I=2}^{4}\gamma_{I}\,^{(I)}Q_{\alpha\beta}\right)
\wedge\vartheta^\alpha\wedge{}^*\!\, T^{\beta} + Q_{\alpha\beta}
\wedge{}^*\!\left(\sum_{I=1}^{4}\beta_{I}\,^{(I)}Q^{\alpha\beta}\right)
\nonumber\\
&&+ \beta_5({}^{(3)}Q_{\alpha\gamma}\wedge\vartheta^{\alpha})\wedge
{}^*({}^{(4)}Q^{\beta\gamma}\wedge\vartheta_{\beta})\Bigg] - \frac{1}{2}
z_4\,R^{\alpha\beta}\wedge{}^*\!{}^{(4)}Z_{\alpha\beta},\label{lagr1}
\end{eqnarray} 
and 
\begin{eqnarray}
&& \alpha_1=a_1 + a_0,\quad \alpha_2=a_2 - 2a_0,\quad 
\alpha_3=a_3 - {a_0\over 2},\label{al}\\
&& \beta_1=b_1 - {a_0\over 4},\quad \beta_2=b_2 + {a_0\over 2},\quad 
\beta_3=b_3 + {a_0\over 8},\quad \beta_4=b_4 - {3a_0\over 8},\quad 
\beta_5=b_5 - a_0,\label{be}\\
&& \gamma_2=c_2 - {a_0\over 2},\quad \gamma_3=c_3 + a_0,\quad  
\gamma_4=c_4 + a_0.\label{ga}
\end{eqnarray}
Correspondingly, the field momenta (\ref{M1})-(\ref{Hab1}) and the
gauge momentum and hypermomentum (\ref{Ea}), (\ref{Eab}) 
can be rewritten in the form
\begin{eqnarray}
&&M^{\alpha\beta}=M^{(0)\alpha\beta} + \widehat{M}^{\alpha\beta},\quad\quad
H_{\alpha}=H^{(0)}_{\alpha} + \widehat{H}_{\alpha}, \quad\quad
H^{\alpha}{}_{\beta}=H^{(0)\alpha}{}_{\beta} + \widehat{H}^{\alpha}{}_{\beta},
\\ &&E_{\alpha}=E^{(0)}_{\alpha} + \widehat{E}_{\alpha}, \quad\quad
E^{\alpha}{}_{\beta}=E^{(0)\alpha}{}_{\beta} + \widehat{E}^{\alpha}{}_{\beta},
\label{Edec}
\end{eqnarray}
where 
\begin{eqnarray}
\widehat{M}^{\alpha\beta}:=-2{\partial\widehat{V}\over\partial 
Q_{\alpha\beta}}&=& -{2\over\kappa}\Bigg[{}^*\! \left(\sum_{I=1}^{4}
\beta_{I}{}^{(I)}Q^{\alpha\beta}\right) + {1\over 2}\beta_5\left(
\vartheta^{(\alpha}\wedge{}^*(Q\wedge\vartheta^{\beta)}) - 
{1\over 4}g^{\alpha\beta}\,{}^*(3Q + \Lambda)\right)\nonumber\\
&& +\,\gamma_{2}\,\vartheta^{(\alpha}\wedge{}^*\! ^{(1)}T^{\beta)} +
\gamma_{3}\,\vartheta^{(\alpha}\wedge{}^*\! ^{(2)}T^{\beta)} +{1\over 4}
(\gamma_{3}-\gamma_{4})\,g^{\alpha\beta}{}^*\!\,T\Bigg]\,,\label{M2}\\
\widehat{H}_{\alpha}:=-{\partial\widehat{V}\over\partial T^{\alpha}}
&=& - {1\over\kappa}\,{}^*\!\left[\left(\sum_{I=1}^{3}\alpha_{I}
{}^{(I)}T_{\alpha}\right) + \left(\sum_{I=2}^{4}\gamma_{I}{}^{(I)}
Q_{\alpha\beta}\wedge\vartheta^{\beta}\right)\right],\label{Ha2}\\
\widehat{H}^{\alpha}{}_{\beta}:= - {\partial\widehat{V}\over\partial
R_{\alpha}{}^{\beta}}&=& z_4\,{}^{*}\!\left({}^{(4)}Z^{\alpha}
{}_{\beta}\right)= {z_4\over 2}\delta^{\alpha}_{\beta}\;{}^*dQ,\label{Hab2}
\end{eqnarray}
and $\widehat{E}_{\alpha}, \widehat{E}^{\alpha}{}_{\beta}$ are defined by
putting ``hats'' over corresponding terms in (\ref{Ea})-(\ref{Eab}).

Taking into account (\ref{Hab2}) and the identities (\ref{ident}), one can
transform the field equations of MAG (\ref{first}) and (\ref{second}) into
\begin{eqnarray} 
{a_0\over 2}\,\widetilde{R}^{\mu\nu}\wedge\eta_{\alpha\mu\nu} 
&=&\kappa\left(\Sigma_{\alpha} - D\widehat{H}_{\alpha} + \widehat{E}_{\alpha}
\right), \label{first1}\\ {z_4\over 2}g_{\alpha\beta}\,d{}^*dQ +
\vartheta_{(\alpha}\wedge\widehat{H}_{\beta)} + \widehat{M}_{\alpha\beta}
&=&\Delta_{(\alpha\beta)},\label{secondS}\\
\vartheta_{[\alpha}\wedge\widehat{H}_{\beta]}&=&\Delta_{[\alpha\beta]}.
\label{secondA}
\end{eqnarray}
The last two equations are clearly the symmetric and the antisymmetric parts
of (\ref{second}). 

Observe that the splitting in (\ref{split})-(\ref{Edec}) has two important 
consequences: With the help of the identities (\ref{ident}), the 
first MAG equation reduces to the Einstein equation with some effective 
source on the right-hand side (\ref{first1}), whereas the gauge field momenta 
(\ref{M2}) and (\ref{Ha2}) are linear combinations of irreducible parts of 
torsion and nonmetricity. Thus, in order to solve the second MAG field 
equation (\ref{secondS})-(\ref{secondA}), we require an irreducible 
decomposition (similar to those established for torsion and nonmetricity in 
section 2) of the gauge field momenta.

\subsection{Irreducible decomposition of ${}^*\widehat{H}_{\alpha}$.}

It turns out that technically it is more convenient not to decompose the
gauge field momentum but rather its Hodge dual, ${}^*\widehat{H}_{\alpha}$. 
This quantity is a vector-valued 2-form, exactly like the torsion form, and
hence its irreducible decomposition has the same structure:
${}^*\widehat{H}_{\alpha}={}^{(1)}({}^*\widehat{H}_{\alpha}) +
{}^{(2)}({}^*\widehat{H}_{\alpha}) + {}^{(3)}({}^*\widehat{H}_{\alpha})$, 
where the three irreducible pieces are defined along the same lines 
as (\ref{T2})-(\ref{T1}). After some algebra, we derive from (\ref{Ha2}) 
the following expressions:
\begin{eqnarray}
{}^{(1)}({}^*\widehat{H}^{\alpha})&=&{1\over\kappa}\left(
\alpha_1\,{}^{(1)}T^{\alpha} - \gamma_2\,{}^*\Omega^{\alpha}\right),
\label{H1}\\
{}^{(2)}({}^*\widehat{H}^{\alpha})&=&{1\over 3\kappa}
\vartheta^{\alpha}\wedge\Big(\alpha_2\, T + \gamma_3\,\Lambda -
3\gamma_4\,Q\Big),\label{H2}\\
{}^{(3)}({}^*\widehat{H}^{\alpha})&=&{1\over\kappa}\alpha_{3}\,
{}^{(3)}T^{\alpha}.\label{H3}
\end{eqnarray}

As is well known (see, e.g., \cite{PR} Appendix A.1.7), an algebraic
equation of the type (\ref{secondA}) can be solved explicitly with respect
to the gauge field momentum. The solution reads
\begin{equation}
\widehat{H}^{\alpha}= - 2e_{\beta}\rfloor\Delta^{[\alpha\beta]} +
{1\over 2}\vartheta^{\alpha}\wedge\left(e_{\mu}\rfloor e_{\nu}\rfloor
\Delta^{[\mu\nu]}\right).\label{Hsource}
\end{equation}

In our case, the spin current 3-form $\Delta^{[\alpha\beta]}$ is given in
terms of the hyperfluid expression (\ref{hcur}). Substituting this into the 
dual of the right-hand side of (\ref{Hsource}), decomposing it into 
irreducible pieces, and using (\ref{H1})-(\ref{H3}) for the dual of the 
left-hand side of (\ref{Hsource}), we find that
\begin{eqnarray}
\alpha_1\,{}^{(1)}T^{\alpha} - \gamma_2\,{}^*\Omega^{\alpha}&=&
{4\kappa\over 3}\,u^{(\alpha}\tau^{\mu)\nu}\,\vartheta_{\mu}\wedge
\vartheta_{\nu},\label{HT1}\\
\alpha_2\, T + \gamma_3\,\Lambda - 3\gamma_4\,Q &=& 0,\label{HT2}\\
\alpha_{3}\,{}^{(3)}T^{\alpha}&=& -\,{\kappa\over 2}\,
u^{[\alpha}\tau^{\mu\nu]}\,\vartheta_{\mu}\wedge\vartheta_{\nu}
,\label{HT3}
\end{eqnarray}
where we have defined $\tau_{\mu\nu}:=J_{[\mu\nu]}$ and used the Frenkel
condition (\ref{frenk}).

\subsection{Irreducible decomposition of ${}^*\widehat{M}_{\alpha\beta}$.}

We observe that the Hodge dual of the gauge momentum,
${}^*\widehat{M}_{\alpha\beta}$, is a symmetric tensor-valued 1-form, exactly
like the nonmetricity $Q_{\alpha\beta}$. Hence, we can decompose this
quantity into four irreducible parts ${}^*\widehat{M}_{\alpha\beta}=
{}^{(1)}({}^*\widehat{M}_{\alpha\beta}) + {}^{(2)}({}^*
\widehat{M}_{\alpha\beta}) + {}^{(3)}({}^*\widehat{M}_{\alpha\beta}) +
{}^{(4)}({}^*\widehat{M}_{\alpha\beta})$, the structure of which
is determined by the pattern (\ref{Q2})-(\ref{Q1}). From (\ref{M2}) we
find
\begin{eqnarray}
{}^{(1)}({}^*\widehat{M}_{\alpha\beta})&=&-\,{2\over\kappa}\beta_1\,
{}^{(1)}Q_{\alpha\beta},\label{Mdec1}\\
{}^{(2)}({}^*\widehat{M}_{\alpha\beta})&=&-\,{2\over\kappa}\,{}^*\left(
\vartheta_{(\alpha}\wedge\left[{2\over 3}\beta_2\,\Omega_{\beta)} +
\gamma_2\,{}^{*(1)}T_{\beta)}\right]\right),\label{Mdec2}\\
{}^{(3)}({}^*\widehat{M}_{\alpha\beta})&=&-\,{2\over\kappa}\left(
\vartheta_{(\alpha}e_{\beta)}\rfloor{\cal M} - 
{1\over 4}g_{\alpha\beta}{\cal M}\right),\label{Mdec3}\\
{}^{(4)}({}^*\widehat{M}_{\alpha\beta})&=&-\,{2\over\kappa}g_{\alpha\beta}
\left(\beta_4\,Q - {1\over 8}\beta_5\,\Lambda - {1\over 4}\gamma_4\,T\right),
\label{Mdec4}
\end{eqnarray}
where the 1-form ${\cal M}$ is defined by
\begin{equation}
{\cal M}:= {4\over 9}\beta_3\,\Lambda + {1\over 3}\gamma_3\, T -
{1\over 2}\beta_5\,Q.\label{mcal}
\end{equation}

Let us analyze the symmetric equation (\ref{secondS}). Separating out the
trace yields
\begin{equation}
z_4\,d\,{}^*dQ +{1\over\kappa}\,{}^*\!\left(-4\beta_4\,Q +
{1\over 2}\beta_5\,\Lambda + \gamma_4\,T\right)={1\over 2}\Delta,\label{max1}
\end{equation}
where $\Delta:=\Delta^{\alpha}{}_{\alpha}$ denotes the dilation current 
3-form. For the hyperfluid we find $\Delta= J^{\alpha}{}_{\alpha}\,u$.
Notice that, in view of the Frenkel condition (\ref{frenk}), 
$\vartheta^{\alpha}\wedge\widehat{H}_{\alpha}=0$ is an immediate consequence
of (\ref{Hsource}). Subtracting (\ref{max1}) from (\ref{secondS})
yields a traceless algebraic equation which relates torsion and 
nonmetricity to the pure shear current ${\nearrow\!\!\!\!\!\!\!\Delta}
_{(\alpha\beta)}:=\Delta_{(\alpha\beta)} - {1\over 4}g_{\alpha\beta}\Delta$.
Substituting (\ref{Hsource}) into (\ref{secondS}) and decomposing the
Hodge dual of the resulting traceless equation, we find, after some algebra 
and on comparison with (\ref{Mdec1})-(\ref{Mdec3}), 
\begin{eqnarray}
\beta_1\,{}^{(1)}Q_{\alpha\beta}&=& -{\kappa\over 2}\left(u_{(\alpha}
\zeta_{\beta\gamma)} - {\zeta\over 6}\,u_{(\alpha}g_{\beta\gamma)}\right)
\vartheta^{\gamma},\label{MS1}\\
{2\over 3}\beta_2\,\Omega_{\alpha} + \gamma_2\,{}^{*(1)}T_{\alpha}&=&
-{\kappa\over 3}\left(u_{\mu}\left[\zeta_{\alpha\nu} - {1\over 3}\,\zeta\,
g_{\alpha\nu}\right] + 2u_{(\alpha}\tau_{\mu)\nu}\right)\eta^{\mu\nu},
\label{MS2}\\ 
{\cal M} &=& {\kappa\over 18}\,{}^*\Delta, \label{MS3}
\end{eqnarray}
where $\zeta_{\alpha\beta}:=J_{(\alpha\beta)}$ is the strain (shear plus 
dilation) density and $\zeta:=\zeta^\alpha{}_\alpha=J^\alpha{}_\alpha$ is 
the pure dilation density.

\section{Generic solutions for torsion and nonmetricity}

We are now in a position to determine the irreducible parts of torsion and
nonmetricity as solutions of the second field equation of MAG which has 
been separated into its symmetric and antisymmetric parts, (\ref{secondS}) and 
(\ref{secondA}), respectively. In order to achieve this, we have to take the 
final step and resolve the combined system of algebraic equations 
(\ref{HT1})-(\ref{HT3}), (\ref{MS1})-(\ref{MS3}). Firstly, let us assume that 
the coupling constants $\alpha_1, \beta_2, \gamma_2$ are such that
\begin{equation}
k_3 := 2\alpha_1\beta_2 -3\gamma_2^2 \neq 0.\label{k3}
\end{equation}
Then (\ref{HT1}) and (\ref{MS2}) yield:
\begin{eqnarray}
{}^{(1)}T_{\alpha}&=&\kappa\left({2({4\over 3}\beta_1 + \gamma_2)\over k_3} 
u_{(\alpha}\tau_{\mu)\nu} + {\gamma_2\over k_3}u_{\mu}\left[\zeta_{\alpha\nu} 
- {1\over 3}\,\zeta\,g_{\alpha\nu}\right]\right)\vartheta^{\mu}
\wedge\vartheta^{\nu},\label{Tsol1}\\
{}^*\Omega_{\alpha}&=& \kappa\left({2(\alpha_1 + 2\gamma_2)\over k_3} 
u_{(\alpha}\tau_{\mu)\nu} + {\alpha_1\over k_3}u_{\mu}\left[\zeta_{\alpha\nu} 
- {1\over 3}\,\zeta\,g_{\alpha\nu}\right]\right)\vartheta^{\mu}
\wedge\vartheta^{\nu}.\label{Osol}
\end{eqnarray}

Next, let us introduce three more constants
\begin{equation}
k_0:=4\alpha_2\beta_3 -3\gamma^2_3,\quad\quad
k_1:=9\left({1\over 2}\alpha_2\beta_5 - \gamma_3\gamma_4\right),\quad\quad
k_2:=3\left(4\beta_3\gamma_4 - {3\over 2}\beta_5\gamma_3\right),\label{3k}
\end{equation}
and assume that $k_0\neq 0$.

Then equations (\ref{HT2}) and (\ref{MS3}), considered as an algebraic 
system for $\Lambda$ and $T$ (recall (\ref{mcal})), yield
\begin{eqnarray}
\Lambda &=& {k_1\over k_0}Q + \kappa{\alpha_2\over 2k_0}\,{}^*\!\Delta,
\label{Lsol}\\
T &=&{k_2\over k_0}Q - \kappa{\gamma_3\over 2k_0}\,{}^*\!\Delta .\label{Tsol}
\end{eqnarray}
Substituting this into (\ref{max1}), we find
\begin{equation}
z_4\left( d\,{}^*\! dQ + m^2\,{}^*\! Q\right)=
{1\over 2}\left(1 - {k_1\over 9k_0}\right)\Delta,\label{proca}
\end{equation}
where we have denoted
\begin{equation}
m^2 := {1\over z_4\kappa}\left(-4\beta_4 + {k_1\over 2k_0}\beta_5 +
{k_2\over k_0}\gamma_4\right).\label{mass}
\end{equation}

Thus, all the post-Riemannian geometrical quantities are now determined.
The Weyl 1-form $Q$ satisfies the {\it Proca-type} differential equation 
(\ref{proca}), which describes a covector particle of mass $m$ interacting 
with the dilation current $\Delta=\zeta\,u$. The remaining irreducible torsion 
and nonmetricity pieces are constructed algebraically from the Weyl covector 
$Q$, the spin current $\Delta_{[\alpha\beta]}=\tau_{\alpha\beta}\,u$ and 
the strain current $\Delta_{(\alpha\beta)}=\zeta_{\alpha\beta}\,u$. 

To summarize, the 1st,
2nd, and 3rd pieces of the torsion are described by the eqs. (\ref{Tsol1}), 
(\ref{Tsol}), and (\ref{HT3}), respectively, whereas the 1st, 2nd, and 3rd  
pieces of the nonmetricity are given by (\ref{MS1}), (\ref{Osol}), and
(\ref{Lsol}), respectively. 

This completes the solution of the second field equation of MAG
(\ref{secondS})-(\ref{secondA}). We now turn to the analysis of the
first field equation of MAG which has the form of an effective Einstein 
equation (\ref{first1}).

\section{Effective Einstein theory}

It is a straightforward task to substitute the results of the previous
section into the right-hand side of (\ref{first1}), but an extremely lengthy 
calculation is required to simplify the result. We have to 
expand the covariant exterior derivatives $D$ (which appear in (\ref{en}) and 
in (\ref{first1})) in terms of the Riemannian operator $\widetilde{D}$ and 
possible contributions from the distortion 1-form (\ref{N}). At this stage 
the decomposition (\ref{N1}) is most useful. Another (related) point is the
substitution of torsion and nonmetricity into the covariant exterior 
derivatives in the equations of motion of the hypermomentum (\ref{hy}), 
which then reduce to
\begin{eqnarray}
\widetilde{D}(\tau_{\alpha\beta}u)&=& - u_{\alpha}u^{\lambda}
\widetilde{D}(\tau_{\lambda\beta}u) + u_{\beta}u^{\lambda}
\widetilde{D}(\tau_{\lambda\alpha}u),\label{spin}\\
\widetilde{D}(\sigma_{\alpha\beta}u)&=& - u_{\alpha}u^{\lambda}
\widetilde{D}(\sigma_{\lambda\beta}u) - u_{\beta}u^{\lambda}
\widetilde{D}(\sigma_{\lambda\alpha}u)\nonumber\\
&& - 2\kappa (A + B) 
\,\tau^{\lambda}{}_{(\alpha}\sigma_{\beta)\lambda}\,\eta,\label{shear}
\end{eqnarray}
where the constants $A, B$ are given below in eqs. (\ref{A})-(\ref{B}).

As a result of this calculation we find that the effective Einstein equation 
(\ref{first1}) reads 
\begin{equation}
{a_0\over 2}\eta_{\alpha\beta\gamma}\wedge\widetilde{R}^{\beta\gamma}
+ \lambda\eta_{\alpha} = \kappa\left(\Sigma_{\alpha}^{\rm fluid} +
\Sigma_{\alpha}^{\rm weyl}\right),\label{ein}
\end{equation}
where the effective energy-momentum currents are given by 
\begin{equation}
\Sigma_{\alpha}^{\rm fluid}:=\varepsilon_{\rm eff}\,u u_{\alpha} + 
p_{\rm eff}(\eta_{\alpha} + uu_{\alpha}) +
\eta^{\beta}\Big\{2(g^{\mu\nu}-u^{\mu}u^{\nu})e_{\mu}\rfloor\widetilde{D}
(u_{(\alpha}\tau_{\beta)\nu}\Big\}.\label{efl}
\end{equation}
\begin{equation}
\Sigma_{\alpha}^{\rm weyl}:={z_4\over 2}\left\{
(e_{\alpha}\rfloor dQ)\wedge{}^{\ast}dQ - (e_\alpha\rfloor
{}^{\ast}dQ)\wedge dQ + m^2\Big[(e_{\alpha}\rfloor Q)^{\ast}\! Q 
+ (e_{\alpha}\rfloor^{\ast}\! Q)\wedge Q\Big]\right\},\label{emax}
\end{equation}
In the derivation of (\ref{ein})-(\ref{efl}), the equation of motion 
(\ref{spin}) of spin was used. The effective energy and pressure are 
defined by
\begin{eqnarray}
\varepsilon_{\rm eff}&=&\varepsilon - {\kappa\over 2}\left(
A\tau_{\mu\nu}\tau^{\mu\nu} + B\zeta_{\mu\nu}\zeta^{\mu\nu}
- C\,\zeta^2\right),\label{efen}\\
p_{\rm eff}&=&p - {\kappa\over 2}\left(A\tau_{\mu\nu}\tau^{\mu\nu} 
+ B\zeta_{\mu\nu}\zeta^{\mu\nu} - C\,\zeta^2\right),\label{efp}
\end{eqnarray}
where we denoted
\begin{eqnarray}
A&=&{3\alpha_1 + 12\gamma_2 + 8\beta_2\over 3k_3} + 
{1\over 6\alpha_3},\label{A}\\
B&=&{\alpha_1\over 3k_3}+{1\over 12\beta_1},\label{B}\\
C&=&{\alpha_1\over 9k_3}+{1\over 72\beta_1}-{\alpha_2\over 36k_0}.\label{C}
\end{eqnarray}

It is straightforward to see from (\ref{spin})-(\ref{shear}) that the
quadratic invariants constructed from spin and strain satisfy
\begin{eqnarray}
u\wedge d(\tau_{\mu\nu}\tau^{\mu\nu})&=&2\tau_{\mu\nu}\tau^{\mu\nu}
\,du,\label{spin2}\\
u\wedge d(\zeta_{\mu\nu}\zeta^{\mu\nu})&=&2\zeta_{\mu\nu}\zeta^{\mu\nu}
\,du.\label{shear2}
\end{eqnarray}
As usual, the translational equations of motion for matter can be obtained
from the (effective) Einstein equation. Since the covariant differential 
$\widetilde{D}$ of the left-hand side of (\ref{ein}) vanishes, one finds
\begin{eqnarray}
\widetilde{D}\left(\Sigma_{\alpha}^{\rm fluid}
+\Sigma_{\alpha}^{\rm weyl}\right) &=&
u_{\alpha}\left[ - u\wedge d\varepsilon_{\rm eff} + 
\left(\varepsilon_{\rm eff} + p_{\rm eff}\right)du\right]\nonumber\\
&&-\left(\varepsilon_{\rm eff} + p_{\rm eff}\right)u\wedge\widetilde{D}
u_{\alpha} - (\eta_{\alpha} + uu_{\alpha})dp_{\rm eff}\nonumber\\
&& - (e_{\alpha}\rfloor\widetilde{R}_{\mu\nu})\wedge\Delta^{[\mu\nu]}
- {1\over 2}\left(1 - {k_1\over 9k_0}\right)(e_{\alpha}\rfloor dQ)
\wedge\Delta\nonumber\\
&&+ 2\widetilde{D}\left(\Delta_{[\alpha\beta]}u^{\lambda}e_{\lambda}\rfloor
\widetilde{D}u^{\beta}\right) =0,\label{trans}
\end{eqnarray}
where (\ref{proca}), (\ref{spin}), and the Ricci identity 
$\widetilde{R}_{\mu\nu}\wedge\vartheta^{\nu}=0$ were used. Contracting
(\ref{trans}) with $u^{\alpha}$ and using (\ref{spin2})-(\ref{shear2}),
one recovers the standard continuity equation
\begin{equation}
u\wedge d\varepsilon - (\varepsilon + p)du =0.\label{conteq}
\end{equation}

\section{General vacuum solution}

The main aim of this paper is to consider the hyperfluid as a specific 
example for a material source of MAG. However, it is also interesting to 
study the {\it vacuum case} of the MAG model (\ref{lagr}), which is 
recovered by putting all the material variables equal to zero, $\varepsilon
=p=\tau_{\alpha\beta}=\zeta_{\alpha\beta}=0$. In this case, our decomposition 
analysis provides us with the exact general vacuum solution for the 
post-Riemannian pieces. Namely, it follows from the equations 
(\ref{HT1})-(\ref{HT3}) and (\ref{MS1})-(\ref{MS3}) that, in the 
{\it generic case},
\begin{equation}
\alpha_3\neq 0, \quad \beta_1\neq 0, \quad k_0\neq 0,
\quad k_3\neq 0,\label{gen}
\end{equation}
(see eq. (\ref{k3})) the general solution for torsion and nonmetricity
reads
\begin{eqnarray}
&&{}^{(1)}T^{\alpha}={}^{(3)}T^{\alpha}=0,\quad\quad
{}^{(1)}Q_{\alpha\beta}={}^{(2)}Q_{\alpha\beta}=0,\label{TQ0}\\
&&Q=k_0\phi,\quad \Lambda=k_1\phi, \quad T=k_2\phi,\label{triplet}
\end{eqnarray}
where $\phi$ is a 1-form. We have used (\ref{Lsol})-(\ref{Tsol}) to derive 
the last line. Substituting this into (\ref{proca}) and (\ref{emax}), we
are left with the Einstein-Proca system of equations for the
metric and the $\phi$ field,
\begin{eqnarray}
{a_0\over 2}\eta_{\alpha\beta\gamma}\wedge\widetilde{R}^{\beta\gamma}
+ \lambda\eta_{\alpha} &=& \kappa\Sigma_{\alpha}^{\{\phi\}},\\ 
d\,{}^*\!d\phi + m^2\,{}^*\!\phi &=& 0,\label{proc}
\end{eqnarray}
where $\Sigma_{\alpha}^{\{\phi\}}={1\over 2}z_4 k_0^2\left\{
(e_{\alpha}\rfloor d\phi)\wedge{}^{\ast}d\phi - (e_{\alpha}\rfloor
{}^{\ast}d\phi)\wedge d\phi + m^2\left[(e_{\alpha}\rfloor \phi)^{\ast}\!\phi 
+ (e_{\alpha}\rfloor^{\ast}\!\phi)\wedge\phi\right]\right\}$. Using the
codifferential $\delta$ and the Laplace-Beltrami operator $\Box := d\delta +
\delta d$, one can rewrite (\ref{proc}) in the equivalent form
\begin{equation}
(\Box + m^2)\phi =0,\quad\quad \delta\phi=0.
\end{equation}

The {\it 1-form triplet} (\ref{triplet}), first discovered in
\cite{magexact,magkerr}, was shown to yield the effective Einstein-Proca
system in \cite{tuck}. We have now obtained a much stronger result: 
(\ref{TQ0})-(\ref{triplet}) is not merely a convenient ansatz which describes 
a particular vacuum solution of the MAG model (\ref{lagr}), but is, in fact, 
its unique and the most general vacuum solution. 

For some special choices of the coupling constants, the condition 
(\ref{gen}) may be violated; in \cite{tuck}, e.g.,  the special case 
$\alpha_3=0$ was considered. Then {\it in vacuum}, as was noticed in 
\cite{tuck}, eq. (\ref{HT3}) allows for an arbitrary 3rd irreducible torsion
piece, ${}^{(3)}T^{\alpha}$ (or, equivalently, the pseudotrace $P$ 1-form). 
However, such degenerate special MAG models are clearly unphysical, because, 
{\it in the presence of matter}, an unacceptable constraint will be imposed 
on the source. The above mentioned $\alpha_3=0$ yields, via (\ref{HT3}), the 
vanishing of the spin current, $\Delta_{[\alpha\beta]}=\tau_{\alpha\beta}u=0$. 
Hence, we should confine our attention to the generic models satisfying 
(\ref{gen}), and we discard the non-generic cases as unphysical. [In 
this way, one avoids unphysical solutions with free functions, which is a 
well-known problem in the double duality approach to Poincar\'e gravity 
\cite{dda}].

\section{Cosmology with hyperfluid}

As an example of non-vacuum dynamics of MAG, let us consider a cosmological
model with a hyperfluid as material source. As is well known, the 
hydrodynamical description of cosmological matter is considered as a 
reasonable approximation to a realistic physical source both in the early and 
in the final stages of the universe's evolution. While the cosmology in 
Einstein's general relativity is confined to an ideal fluid with structureless 
elements, in MAG the hyperfluid represents a less trivial medium with 
microstructure, see \cite{capr}.

Before starting the discussion, let us specialize our general model 
(\ref{lagr}) a bit. Although the Lagrangian (\ref{lagr}) involves 11 coupling 
constants $(a_{I}, b_{J}, c_{K})$, they can be combined, as we have seen, 
into only four essential parameters, $m^2, A, B, C$, which
completely determine the dynamics of the effective Einstein-Proca-hyperfluid 
system. Hence there is some freedom in the choice of the coupling constants 
without basically changing the physical content of the model. In this
section we will make use of this freedom in order to study more closely 
the model which has attracted most attention in the literature, see
\cite{Palatini,Pono,TW,gh,magexact,magkerr,tuck}. Consequently, let us 
specialize to the case 
\begin{equation}
a_{I}=0,\; {\scriptstyle I}=1,2,3,\quad b_{J}=0,\; {\scriptstyle J}=1,2,3,5, 
\quad c_{K}=0,\; {\scriptstyle K}=2,3,4,
\label{choice}
\end{equation}
so that only $b_4\neq 0$. Then the  Lagrangian (\ref{lagr}) reduces to a
more manageable form
\begin{equation}
V_{\rm dil} =
\frac{1}{2\kappa}\,\left( -a_0\,R^{\alpha\beta}\wedge\eta_{\alpha\beta}
+ 4b_4 Q\wedge{}^*Q\right) - \frac{1}{2}
z_4\,R^{\alpha\beta}\wedge{}^*{}^{(4)}Z_{\alpha\beta}.\label{dil}
\end{equation}
Substituting (\ref{choice}) into (\ref{al})-(\ref{ga}), (\ref{k3}), (\ref{3k}),
(\ref{mass}), (\ref{A})-(\ref{C}), we find
\begin{eqnarray}
&& k_0=-4a_0^2,\quad k_1=0,\quad k_2=6a_0^2,\quad k_3={1\over 4}a_0^2,\\
&& m^2=-{4b_4\over z_4\kappa},\quad A=B={1\over a_0},\quad C={3\over 8a_0}.
\label{ABC}
\end{eqnarray}

As we can see from Sect.~7, the gravitationally interacting hyperfluid in 
the MAG model (\ref{lagr}) produces an effect similar to that of matter with 
spin \cite{PR} in the usual Einstein-Cartan theory: The total hypermomentum 
density contributes quadratic terms which modify the energy and pressure 
according to (\ref{efen})-(\ref{efp}). Assuming the absence of the strain
current, we recover the Einstein-Cartan theory interacting with a Proca-like
Weyl covector $Q$. The dilation density $\zeta$ ``counteracts'' the 
spin and shear, both of which produce an effective repulsion. The resulting
dynamics of the gravitational field depends crucially on the relative 
values of the quadratic terms in (\ref{efen})-(\ref{efp}). 

Since the effect of pure spin (effective repulsion) is well known 
in cosmology, let us concentrate here on the particular case of hyperfluid 
with diagonal specific hypermomentum density, namely $\mu^{A}{}_{B}=\mu\,
\delta^{A}_{B}$. Then (\ref{mat}) reduces to the dilation hyperfluid, the 
elements of which have only one ``internal'' degree of freedom: they can 
uniformly (in an element's rest frame) change their scale. Examples of such 
media are well known in non-relativistic continuum mechanics. These are, e.g., 
continua with finely dispersed spherical voids and liquids with non-diffusing 
gas bubbles \cite{capr}. The hypermomentum current (\ref{hcur}) is then 
determined by the hypermomentum density 
\begin{equation}
J^{\alpha}{}_{\beta}=\zeta^{\alpha}{}_{\beta}=
{1\over 3}(\delta^{\alpha}_{\beta} + u^{\alpha}u_{\beta})\zeta,
\quad\quad \tau^{\alpha}{}_{\beta}=0,\label{dilcur}
\end{equation}
so that the effective term in the energy and pressure (recall (\ref{ABC})) 
reads
\begin{equation}
{1\over 2a_0}\left(\zeta_{\mu\nu}\zeta^{\mu\nu}- {3\over 8}\zeta^2\right)
=-\,{1\over 48a_0}\zeta^2.\label{corr}
\end{equation}
Therefore we conclude that purely dilational matter amplifies gravitational
attraction. In particular, it accelerates rather than retards the possible
collapse of a system. This happens, though, at extremely small distances
due to the smallness of the correction (\ref{corr}) which enters 
(\ref{efen})-(\ref{efp}) with the gravitational constant $\kappa$. 

In the general case, a massive dilation (or Weyl) field affects gravitation
in a nontrivial way. However, in homogeneous cosmology, there are solutions 
with $R_{\alpha}{}^{\alpha}=2dQ=0$. In that case the kinetic terms of the
type $(dQ)^2$ in the effective Einstein equation (\ref{emax}) disappear, 
whereas pure mass terms $(Q)^2$ simply supply new corrections to energy 
and pressure. Let us be more specific and look for the standard cosmological 
solutions with the space-time interval in the Friedman form,
\begin{equation}
ds^2 = - dt^2 + R^2(t)\left({dr^2\over{1-Kr^2}} + r^2 d\theta^2 +
r^2 \sin^2\theta d\phi^2\right).\label{metr}
\end{equation}
Substituting (\ref{metr}) into the effective Einstein equations (\ref{ein})
and taking (\ref{shear2}) into account, we find
\begin{eqnarray}
3\left({\dot{R}^2\over R^2} + {K\over R^2}\right)&=&\kappa\left\{\varepsilon +
{\kappa\over 48a_0}\left(1-{3a_0\over b_4}\right){\zeta^{2}_{0}\over R^6}
\right\},\label{eq1}\\ 
-2{{\buildrel .. \over R}\over R} - {\dot{R}^2\over R^2} - {K\over R^2}&=&
\kappa\left\{p + {\kappa\over 48a_0}\left(1-{3a_0\over b_4}\right)
{\zeta^{2}_{0}\over R^6}\right\},\label{eq2}
\end{eqnarray}
where $\zeta_{0}$ is an integration constant. In accordance with (\ref{proca}),
(\ref{hy}), (\ref{shear}), we obtain
\begin{equation}
Q=-{\kappa \zeta_{0}\over 8b_4}{dt\over R^3(t)}.
\end{equation}
Evidently $dQ=0$. 

Supplementing (\ref{eq1})-(\ref{eq2}) by the equation of state 
$p=p(\varepsilon)$, we can solve (\ref{conteq}) explicitly. Let us consider
the case $p=\gamma\,\varepsilon$ with constant $\gamma$. Then (\ref{conteq}) 
yields
\begin{equation}
\varepsilon = {\varepsilon_0\over R^{3(1+\gamma)}},\label{ensol}
\end{equation}
where $\varepsilon_0$ is a positive integration constant. Equation (\ref{eq2})
is redundant, as follows from (\ref{eq1}) and (\ref{ensol}). Thus the 
dynamics of the scale factor $R(t)$ is determined by the first order equation
(\ref{eq1}), with (\ref{ensol}) inserted. Interestingly, for the coupling 
constant $b_4=3a_0$, this dynamics turns out to be completely standard, 
yielding the well-known cosmological solutions of general relativity 
theory. However, if one wants to interpret the $(Q)^2$ term in (\ref{lagr}) 
as the mass term for the dilation field, then one must take a negative $b_4$, 
see (\ref{ABC}). Consequently, the dilation correction $\sim{1\over R^6}$ 
enters into the right-hand side of (\ref{eq1}) with a positive coefficient, 
which corresponds to an additional effective {\it attractive} force dominating 
during the very early stages of evolution. Near the
singularity
\begin{equation}
R^3(t)\approx\left({\kappa\zeta_0\over 4}\,\sqrt{{1\over a_0} - 
{3\over b_4}}\right)\,t.\label{R}
\end{equation}
This is true for any value of the spatial curvature $K$ and for an 
arbitrary equation of state with $0\leq\gamma <1$. 

\section{Conclusion}

In this paper we have applied the irreducible decomposition technique 
to the study of the classical MAG model (\ref{lagr}) which has recently 
attracted quite some attention in the literature. Our main observations 
are as follows: 

Torsion and traceless nonmetricity are explicitly expressible in terms of
the spin and shear currents of the hyperfluid. This enables us to reduce
the general MAG field equations to the effective Einstein theory (\ref{ein})
with a source represented by the energy-momentum tensors of the Weyl 
(Proca-type) covector field (\ref{emax}) and of the effective 
(Weyssenhoff-type) spin fluid (\ref{efl}). 

{\it In vacuum}, the 1-form triplet (\ref{triplet}) describes the general 
and unique solution of the field equations of MAG. This result completes 
previous studies of the 1-form triplet \cite{magexact,magkerr,tuck}.

As an example of a nontrivial case with matter, we have studied homogeneous 
cosmologies with hyperfluid. Like in the Einstein-Cartan theory, we conclude 
that the hypermomentum affects significantly the cosmological evolution only 
in the very early stages. However, contrary to the effect of spin,
shear does not prevent the formation of a cosmological singularity but
rather promotes it. Homogeneous cosmologies in MAG models with {\it 
ideal} fluid were recently studied in \cite{mink}.

\acknowledgments

For YNO this research was supported by the Deutsche Forschungsgemeinschaft 
(Bonn) under project He-528/17-2.

\bigskip

\end{document}